\documentclass[prx,twocolumn,superscriptaddress]{revtex4-1}
\usepackage{amsmath,amssymb,bm,mathrsfs,graphicx, braket,amsthm,enumerate,times}
\usepackage[colorlinks=true,citecolor=blue,linkcolor=blue]{hyperref}

\begin{document}
\title{A proof of the Bloch theorem for lattice models}
\author{Haruki Watanabe} 
\email{haruki.watanabe@ap.t.u-tokyo.ac.jp}
\affiliation{Department of Applied Physics, University of Tokyo, Tokyo 113-8656, Japan}
\begin{abstract}
The Bloch theorem is a powerful theorem stating that the expectation value of the U(1) current operator averaged over the entire space vanishes in large quantum systems.   
The theorem applies to the ground state and to the thermal equilibrium at a finite temperature, irrespective of the details of the Hamiltonian as far as all terms in the Hamiltonian are finite ranged. In this work we present a simple yet rigorous proof for general lattice models.
For large but finite systems, we find that both the discussion and the conclusion are sensitive to the boundary condition one assumes: under the periodic boundary condition, one can only prove that the current expectation value is inversely proportional to the linear dimension of the system, while the current expectation value completely vanishes before taking the thermodynamic limit when the open boundary condition is imposed.
We also provide simple tight-binding models that clarify the limitation of the theorem in dimensions higher than one.
\end{abstract}


\maketitle

\section{Introduction}
The Bloch theorem~\cite{PhysRev.75.502} states that the equilibrium state of a thermodynamically large system, in general, does not support non-vanishing expectation value of the averaged current density of any conserved U(1) charge, regardless of the details of the Hamiltonian such as the form of interactions or the size of the excitation gap.  Despite its wide applications, the proof of the theorem in the existing literature is mostly for specific continuum models~\cite{PhysRev.75.502,doi:10.1143/JPSJ.65.3254,PhysRevD.92.085011}. There are also proofs for lattice models~\cite{PhysRevB.78.144404,Tada2016,Bachmann} but the setting considered in these works are not fully general. For example, Ref.~\cite{PhysRevB.78.144404} is for a concrete spin model with a translation symmetry and the assumption of their discussion is unclear. Ref.~\cite{Tada2016} assumes an extended-Hubbard type Hamiltonian and their definition of the current operator heavily relies on the specific form of the kinetic term. Finally, Ref.~\cite{Bachmann} assumes the uniqueness of the ground state with non-vanishing excitation gap.

In this work, we revisit the proof and clarify several confusing points about the Bloch theorem.  
We summarize the assumption and the statement of the theorem under the periodic boundary condition in Sec.~\ref{sec:setup} and give a proof for general models defined on a one-dimensional lattice in Sec.~\ref{sec:proof}. 
We discuss the theorem under the open boundary condition in Sec.~\ref{sec:open}.
Finally we clarify the limitation of the theorem in higher dimensions in Sec.~\ref{sec:highd}.

\section{Bloch theorem under periodic boundary condition}
\subsection{Setup and statement}
\label{sec:setup}
Let us consider a quantum many-body system defined on a one dimensional lattice.  We impose the periodic boundary condition with system size $L$.  The Hamiltonian $\hat{H}$ of the system can be very general. It may contain arbitrary hopping matrices and interactions as far as each term in the Hamiltonian is short-ranged (i.e., the size of its support is finite and does not scale with $L$) and respects the U(1) symmetry we discuss shortly. In particular, we \emph{do not} put any restriction on the translation symmetry, the ground state degeneracy, or the excitation gap. Hence the result is applicable not only to periodic lattice with arbitrary number of sub-lattices but also, for example, to quasi-crystals or disordered systems.
To simplify the notation we set the lattice constant to be $1$ and denote lattice sites by $x\in\{1,2,\cdots,L\}$.

We assume that the Hamiltonian $\hat{H}$ commutes with the particle number operator 
\begin{equation}
\hat{N}=\sum_{x=1}^L\hat{n}_x.
\end{equation}
Here, $\hat{n}_x$ is the local charge density operator at site $x$. We assume that density operators at different sites commute, $[\hat{n}_x,\hat{n}_{x'}]=0$.  The U(1) symmetry implies the conservation law:
\begin{equation}
i[\hat{H},\hat{n}_x]+\hat{j}_{x+\frac{1}{2}}-\hat{j}_{x-\frac{1}{2}}=0,\label{eq:cons}
\end{equation}
where $\hat{j}_{x+\frac{1}{2}}$ is the local U(1) current operator that measures the net charge transfer across the `seam' in between $x$ and $x+1$ [Fig.~\ref{fig} (a)]. We present the precise definition of $\hat{j}_{x+\frac{1}{2}}$ in Sec.~\ref{subsec:current}. 

With this setting, the Bloch theorem states that the ground state expectation value of the local current operator vanishes in the limit of the large system size:
\begin{equation}
\lim_{L\rightarrow\infty}\langle\text{GS}|\hat{j}_{x+\frac{1}{2}}|\text{GS}\rangle=0.\label{statement1}
\end{equation}
Here $|\text{GS}\rangle$ is the ground state of $\hat{H}$ with the energy eigenvalue $E_{\text{GS}}$. When there is a ground state degeneracy we arbitrary pick one state.  The current conservation law in Eq.~\eqref{eq:cons}, together with $\hat{H}|\text{GS}\rangle=E_{\text{GS}}|\text{GS}\rangle$, implies
\begin{equation}
\langle\text{GS}|\hat{j}_{x+\frac{1}{2}}|\text{GS}\rangle=\langle\text{GS}|\hat{j}_{x-\frac{1}{2}}|\text{GS}\rangle\label{constantj}
\end{equation}
for all $x=1,2,\cdots,L$, meaning that the expectation value is independent of the position. Therefore, we can equally state the Bloch theorem in terms of the averaged current operator
\begin{align}
&\hat{\bar{j}}\equiv\frac{1}{L}\sum_{x=1}^L\hat{j}_{x+\frac{1}{2}},\label{eq:ac}\\
&\lim_{L\rightarrow\infty}\langle\text{GS}|\hat{\bar{j}}|\text{GS}\rangle=0.\label{statement12}
\end{align}
This statement can be directly generalized to a finite temperature $T>0$~\cite{doi:10.1143/JPSJ.65.3254,Tada2016} described by the Gibbs state (we set $k_B=1$):
\begin{align}
&\hat{\rho}_0\equiv \frac{1}{Z}e^{-\hat{H}/T},\quad Z\equiv\text{tr}\big(e^{-\hat{H}/T}\big),\label{Gibbs}\\
&\lim_{L\rightarrow\infty}\text{tr}\big(\hat{\rho}_0\hat{j}_{x+\frac{1}{2}}\big)=\lim_{L\rightarrow\infty}\text{tr}\big(\hat{\rho}_0\hat{\bar{j}}\big)=0.\label{statement2}
\end{align}

\subsection{Proof}
\label{sec:proof}
\subsubsection{Variational principle}
\label{variation}
Our proof of the theorem makes use of the twist operator introduced by Ref.~\cite{Lieb1961}, which reads
\begin{equation}
\hat{U}_m\equiv e^{\frac{2\pi im}{L}\sum_{x=1}^Lx\hat{n}_x},\quad m\in\mathbb{Z}.
\label{twistop}
\end{equation}
This unitary operator is consistent with the periodic boundary condition since replacing $x$ with $x+L$ in the exponent does not affect $\hat{U}_m$ as $e^{2\pi im\hat{N}}=1$. The key observation of the proof is the following Taylor expansion in the power series of $L^{-1}$, which we show in Sec.~\ref{subsec:current}:
\begin{equation}
\hat{U}_m^\dagger\hat{H}\hat{U}_m=\hat{H}+2\pi m \hat{\bar{j}}+O(L^{-1}).\label{eq:Taylor}
\end{equation}
Taking the ground state expectation value of this equation, we find the following relation for the energy expectation value of the variational state $|\Phi_m\rangle\equiv\hat{U}_m|\text{GS}\rangle$:
\begin{equation}
\langle\Phi_m|\hat{H}|\Phi_m\rangle=E_{\text{GS}}+2\pi m \langle \text{GS}|\hat{\bar{j}}|\text{GS}\rangle+O(L^{-1}).
\end{equation}
Suppose first that $\langle \text{GS}|\hat{\bar{j}}|\text{GS}\rangle>0$. Then we find that $\langle\Phi_m|\hat{H}|\Phi_m\rangle$ with $m<0$ is lower than the ground state energy for a large $L$, which contradicts with the variational principle.  If $\langle \text{GS}|\hat{\bar{j}}|\text{GS}\rangle<0$, $|\Phi_m\rangle$ with $m>0$ does the same job. Hence, $\langle \text{GS}|\hat{\bar{j}}|\text{GS}\rangle$ cannot remain nonzero as $L\rightarrow\infty$ and must be smaller than or equal to $O(L^{-1})$. The is the proof of Eq.~\eqref{statement12}, which also gives Eq.~\eqref{statement1} with the help of Eq.~\eqref{constantj}.  The above variational argument is common among the majority of proofs in the literature~\cite{PhysRev.75.502,PhysRev.75.502,doi:10.1143/JPSJ.65.3254,PhysRevD.92.085011,PhysRevB.78.144404}.

\subsubsection{Finite temperature}
The proof of the Bloch theorem for a finite temperature is almost identical to that for the ground state. Given the Hamiltonian $\hat{H}$ and a density operator $\hat{\rho}$, in general, the free energy at $T>0$ is given by
\begin{equation}
F(\hat{\rho})=\text{tr}\big(\hat{\rho}\hat{H}+T\hat{\rho}\ln \hat{\rho}\big).
\end{equation}
This is minimized by the Gibbs state in Eq.~\eqref{Gibbs} with the minimum value $F(\hat{\rho}_0)=-k_BT\ln Z$~\cite{PhysRev.106.620,Sagawa}.  
Using Eq.~\eqref{eq:Taylor}, we find
\begin{align}
F(\hat{U}_m\hat{\rho}_0\hat{U}_m^\dagger)&=\text{tr}\left[\hat{\rho}_0(\hat{U}_m^\dagger\hat{H}\hat{U}_m)+T\hat{\rho}_0\ln \hat{\rho}_0\right]\notag\\
&=F(\hat{\rho}_0)-2\pi m\,\text{tr}\big(\hat{\rho}_0\hat{\bar{j}}\big)+O(L^{-1}).
\end{align}
If the magnitude of the current expectation value is bigger than $O(L^{-1})$, we get $F(\hat{\rho}_m)<F(\hat{\rho}_0)$ for either $m=\pm1$, which is a contradiction.

\subsubsection{Local current operator}
\label{subsec:current}

\begin{figure}[t]
\includegraphics[width=1.0\columnwidth]{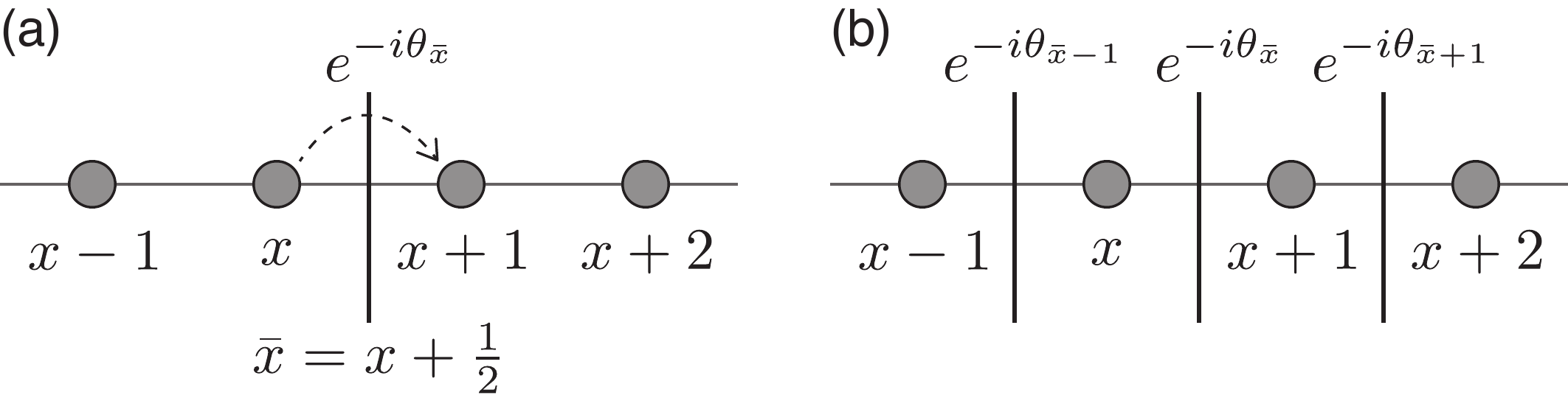}
\caption{\label{fig} (a) Twisted boundary condition with the U(1) phase $e^{-i\theta_{\bar{x}}}$ at the seam $\bar{x}=x+\frac{1}{2}$. (b) Generalized twisted boundary condition with a seam at every $\bar{x}$ for $x=1,2,\cdots,L$.
}\end{figure}

It remains to verify Eq.~\eqref{eq:Taylor}. This requires a precise formulation of the local current operator.  To this end, let us temporary introduce the \emph{twisted} boundary condition. We place the position of the `seam' to be somewhere in between $x$ and $x+1$, which we denote by $\bar{x}\equiv x+\frac{1}{2}$ [Fig.~\ref{fig} (a)].  Let $\theta_{\bar{x}}$ be the angle of the twist. Later we will set $\theta_{\bar{x}}=0$, as, after all, we are interested in the original system under the periodic boundary condition.  

The Hamiltonian $\hat{H}^{\theta_{\bar{x}}}$ under the twisted boundary condition has $\theta_{\bar{x}}$-dependence localized around the seam. This is because every term in the original Hamiltonian $\hat{H}$ that goes across the seam acquires a phase $e^{i\ell_{\bar{x}}\theta_{\bar{x}}}$. For example, a hopping term $tc_{x+1}^\dagger c_x+\text{h.c.}$ becomes $te^{-i\theta_{\bar{x}}}c_{x+1}^\dagger c_x+\text{h.c.}$, while a pair-hopping term $t\hat{c}_{x+1,\uparrow}^\dagger\hat{c}_{x+1,\downarrow}^\dagger\hat{c}_{x,\downarrow}\hat{c}_{x,\uparrow}+\text{h.c.}$ becomes $te^{-2i\theta_{\bar{x}}}\hat{c}_{x+1,\uparrow}^\dagger\hat{c}_{x+1,\downarrow}^\dagger\hat{c}_{x,\downarrow}\hat{c}_{x,\uparrow}+\text{h.c.}$  More generally, a term in $\hat{H}$ is multiplied by the factor $e^{i(n_{\bar{x}}^a-n_{\bar{x}}^c)\theta_{\bar{x}}}$ where $n_{\bar{x}}^a$ ($n_{\bar{x}}^c$) is the number of annihilation (creation) operators in the right side of the seam in the term. Other terms in $\hat{H}$ that reside either one side of the seam remain unchanged.  The local current operator across the seam under the \emph{periodic} boundary condition is given by
\begin{equation}
\hat{j}_{\bar{x}}\equiv\partial_{\theta_{\bar{x}}}\hat{H}^{\theta_{\bar{x}}}\big|_{\theta_{\bar{x}}=0}.\label{eq:lc}
\end{equation}

The current operator defined this way satisfies the conservation law in Eq.~\eqref{eq:cons}. To see this explicitly, let us introduce a seam for every $\bar{x}=x+\frac{1}{2}$ ($x=1,2,\cdots,L$) and denote the twisted Hamiltonian by $\hat{H}^{(\theta_{\bar{1}},\theta_{\bar{2}},\cdots,\theta_{\bar{L}})}$ [Fig.~\ref{fig} (b)]. It satisfies
\begin{align}
&\hat{H}=\hat{H}^{(\theta_{\bar{1}},\theta_{\bar{2}},\cdots,\theta_{\bar{L}})}\big|_{\theta_{\bar{1}}=\theta_{\bar{2}}=\cdots=\theta_{\bar{L}}=0},\\
&\hat{j}_{\bar{x}}=\partial_{\theta_{\bar{x}}}\hat{H}^{(\theta_{\bar{1}},\theta_{\bar{2}},\cdots,\theta_{\bar{L}})}\big|_{\theta_{\bar{1}}=\theta_{\bar{2}}=\cdots=\theta_{\bar{L}}=0}
\end{align}
and
\begin{align}
&e^{i\epsilon \hat{n}_{x}}\hat{H}^{(\theta_{\bar{1}},\theta_{\bar{2}},\cdots,\theta_{\bar{L}})}e^{-i\epsilon \hat{n}_{x}}\notag\\
&=\hat{H}^{(\theta_{\bar{1}},\cdots,\theta_{\bar{x}-2},\theta_{\bar{x}-1}-\epsilon,\theta_{\bar{x}}+\epsilon,\theta_{\bar{x}+1},\cdots,\theta_{\bar{L}})}.\label{eq:Htr}
\end{align}
This relation implies that $\theta_{\bar{x}}$ can be identified with the background U(1) gauge field $A_{\bar{x}}$.  When Eq.~\eqref{eq:Htr} for $\theta_{\bar{1}}=\theta_{\bar{2}}=\cdots=\theta_{\bar{L}}=0$ is expanded in the power series of $\epsilon$, the $O(\epsilon)$-term reproduces the conservation law~\eqref{eq:cons}.  It also follows by using Eq.~\eqref{eq:Htr} repeatedly that
\begin{equation}
\hat{U}_m^\dagger\hat{H}\hat{U}_m=\hat{H}^{(\frac{2\pi m}{L},\cdots,\frac{2\pi m}{L})}.\label{v1}
\end{equation}
The Taylor series of the right-hand side reads
\begin{equation}
\hat{H}^{(\frac{2\pi m}{L},\cdots,\frac{2\pi m}{L})}=\sum_{\ell=0}^\infty\tfrac{1}{\ell!}(\tfrac{2\pi m}{L})^\ell\hat{H}^{(\ell)},
\end{equation}
where $\hat{H}^{(\ell)}$ ($\ell=0,1,2,\cdots$) is defined by
\begin{equation}
\sum_{x_1,x_2,\cdots,x_\ell=1}^L\partial_{\theta_{\bar{x}_1}}\partial_{\theta_{\bar{x}_2}}\cdots\partial_{\theta_{\bar{x}_\ell}}\hat{H}^{(\theta_{\bar{1}},\cdots,\theta_{\bar{L}})}\Big|_{\theta_{\bar{1}}=\cdots=\theta_{\bar{L}}=0}.\label{v2}
\end{equation}
For example, $\hat{H}^{(0)}=\hat{H}$ and
\begin{equation}
\hat{H}^{(1)}=\sum_{x=1}^L\partial_{\theta_{\bar{x}}}\hat{H}^{(\theta_{\bar{1}},\cdots,\theta_{\bar{L}})}\Big|_{\theta_{\bar{1}}=\cdots=\theta_{\bar{L}}=0}=L\hat{\bar{j}}.\label{v3}
\end{equation}
For short-ranged Hamiltonians, each $\hat{H}^{(\ell)}$ is at most the order of $L$ at least for $\ell=O(L^0)$. Eqs.~\eqref{v1}--\eqref{v3} altogether verify Eq.~\eqref{eq:Taylor} and the proof is completed.

\subsection{Discussions}
\label{sec:extention}
Here let us make some remarks on the theorem.
\subsubsection{Relation to the Lieb-Schultz-Mattis theorem}
The conclusion in Sec.~\ref{variation} implies that the variational state $|\Phi_m\rangle=\hat{U}_m|\text{GS}\rangle$ is a low-energy state whose excitation energy $|\langle\Phi_m|\hat{H}|\Phi_m\rangle-E_{\text{GS}}|$ is bounded by $O(L^{-1})$. 
Further assuming the translation symmetry $\hat{T}_1$ with $\hat{T}_1\hat{n}_x\hat{T}_1^\dagger=\hat{n}_{x+1}$, we find~\cite{Lieb1961,Affleck1986,PhysRevLett.79.1110}
\begin{equation}
\hat{T}_1\hat{U}_m\hat{T}_1^\dagger=\hat{U}_me^{-2\pi m i\frac{\hat{N}}{L}}.
\end{equation}
Suppose that the ground state $|\text{GS}\rangle$ is an eigenstate of $\hat{T}_1$ and $\hat{N}$ and that the filling fraction $\nu\equiv \langle\text{GS}|\hat{N}|\text{GS}\rangle/L$ is not an integer. Then the variational state $|\Phi_m\rangle$ and the ground state $|\text{GS}\rangle$ have inequivalent eigenvalues of $\hat{T}_1$ and hence are orthogonal to each other. This implies the Lieb-Schultz-Mattis theorem for translation invariant one-dimensional systems which suggests the presence of either a ground state degeneracy or a gapless excitation when $\nu\notin\mathbb{Z}$~\cite{Lieb1961,Affleck1986,PhysRevLett.79.1110}. Note that we did not assume any additional symmetry such as the spatial inversion or the time-reversal symmetry unlike the original argument~\cite{Lieb1961,Affleck1986,PhysRevLett.79.1110}.
In the context of the Lieb-Schultz-Mattis theorem, it appears that Ref.~\cite{Koma2000} removed such an assumption for first time by using the variational argument, but as we have seen here this logic was old and can be traced back to the original work of the Bloch theorem~\cite{PhysRev.75.502}.

\begin{figure}[t]
\includegraphics[width=1.0\columnwidth]{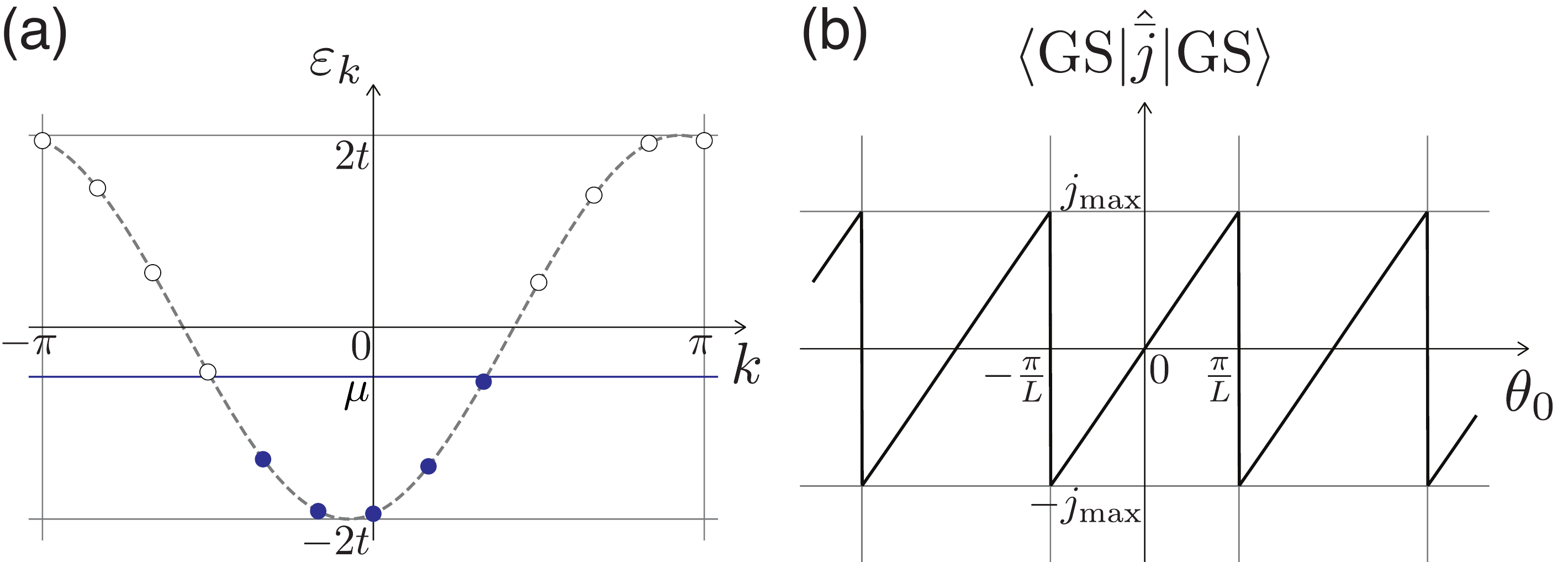}
\caption{\label{fig2} (a) The band structure of the tight-binding model for $L=12$ and $\theta_0=0.9\pi/L$.  Blue (white) dots represents occupied (unoccupied) states at $\mu=-2t\cos(\frac{\pi N}{L})$ with $N=5$.
(b) The current expectation value of the ground state as a function of $\theta_0$ for $L=12$ and $N=5$.
}\end{figure}

\subsubsection{Persistent current in a finite system}
\label{subsec:persistent}
The Bloch theorem allows a persistent current of the order $O(L^{-1})$ in a finite system. For a later purpose, let us consider a concrete tight-binding model with the nearest neighbor hopping $t>0$.
\begin{equation}
\hat{H}=-te^{-i\theta_0}\sum_{x=1}^L\hat{c}_{x+1}^\dagger \hat{c}_x+\text{h.c.}
\end{equation}
We introduced a phase $e^{-i\theta_0}$ to break the time-reversal symmetry. Introducing the Fourier transform $\hat{c}_{k}^\dagger=\frac{1}{\sqrt{L}}\sum_{x=1}^L\hat{c}_{x}^\dagger e^{i k x}$ for $k=\frac{2\pi q}{L}$ ($q=1,2,\cdots L$), we find~\cite{PhysRevB.37.6050}
\begin{equation}
\hat{H}=\sum_k \varepsilon_k \hat{c}_k^\dagger \hat{c}_k,\quad\hat{\bar{j}}=\frac{1}{L}\sum_k\partial_k\varepsilon_k \hat{c}_k^\dagger \hat{c}_k
\end{equation}
with $\varepsilon_k=-2t\cos(k+\theta_0)$ [Fig.~\ref{fig2} (a)].

For concreteness, let us set the Fermi energy to be $\mu=-2t\cos(\frac{\pi N}{L})$ for an odd number of particles $N$, and consider the Fermi sea $|\text{GS}\rangle=\prod_{k, \varepsilon_k<\mu}c_k^\dagger|0\rangle$  [Fig.~\ref{fig2} (a)]. The current expectation value $\langle \text{GS}|\hat{\bar{j}}|\text{GS}\rangle=\frac{1}{L}\sum_{k, \varepsilon_k<\mu}\partial_k\varepsilon_k$ shows the periodicity in $\theta_0$ with the period $2\pi/L$  [Fig.~\ref{fig2} (b)]. Its maximum value is given by
\begin{equation}
j_{\text{max}}\equiv\lim_{\theta_0\uparrow\frac{\pi}{L}}\langle \text{GS}|\hat{\bar{j}}|\text{GS}\rangle=\tfrac{2t}{L}\sin(\tfrac{\pi N}{L})=O(L^{-1}).
\label{eq:jmax}
\end{equation}
These results are consistent with the previous studies, for example, in Ref.~\cite{PhysRevB.37.6050}.

\subsubsection{Extensions}
\paragraph{Continuum models}
Continuum models can be treated simply by replacing $\sum_{x=1}^L$ with $\int_0^Ldx$, for example. For continuum models, the Noether theorem provides the definition of the conserved U(1) current. The key relation Eq.~\eqref{eq:Taylor} remains unchanged.

\paragraph{Long-range interactions}
The assumption on the range of hopping matrices and interactions can be slightly relaxed.  
They are not necessarily strictly finite-ranged.
For example, any term $\hat{o}$ that does not depend on $\theta_{\bar{x}}$ (i.e. $\hat{U}_m^\dagger\hat{o}\hat{U}_m=\hat{o}$) can be safely added.  This class includes the density-density interactions such as the Coulomb interaction among electrons. 

Terms with a long-range tail that depend on $\theta_{\bar{x}}$ are also allowed as long as the order estimate of the series expansion in Eq.~\eqref{eq:Taylor} is preserved.
In addition to exponentially decaying terms, power-low decaying terms with a large enough exponent can also be added.  The minimum value of the power depends on the detailed form of the term.
For instance, in the case of the hopping term
\begin{equation}
t\sum_{n=1}^Ln^{-\alpha}\sum_{x=1}^L\hat{c}_{x+n}^\dagger \hat{c}_x+\text{h.c.},
\end{equation}
the power $\alpha$ must be greater than $3$. [When $2<\alpha\leq3$, the Bloch theorem still holds but the upper bound of the current expectation value decays slower than $O(L^{-1})$.] 

\paragraph{Other conserved current}
The argument in this work coherently applies to any conserved current density associated with an internal U(1) symmetry. 
For example, when the $z$-component of the total spin is conserved in a spin model with spin $S$ ($=1/2, 1, 3/2. \cdots$) on each site, we can set $\hat{n}_x=\hat{S}_x^z-S$~\cite{PhysRevLett.78.1984} in our discussion above to prove the absence of the equilibrium spin current. However, our argument is not applicable, for example, to the energy current density as there does not exist the corresponding twist operator. Recently, a completely new argument for the energy current has been developed in Ref.~\cite{Anton}.

\section{Bloch theorem under open boundary condition}
\label{sec:open}
When we impose the open boundary condition instead of the periodic boundary condition, we can actually prove a stronger statement:
\begin{equation}
\langle\text{GS}|\hat{\bar{j}}|\text{GS}\rangle=\text{tr}\big(\hat{\rho}_0\hat{\bar{j}}\big)=0.\label{openbc}
\end{equation}
Unlike the case with the periodic boundary condition, we do not have to take the large $L$ limit.  To see this, note that the position operator 
\begin{equation}
\hat{P}=\frac{1}{L}\sum_{x=1}^Lx\hat{n}_x
\end{equation}
is well defined under the open boundary condition. This operator is the one in the exponent of the twist operator~\eqref{twistop} and is also known as the polarization operator~\cite{PhysRevX.8.021065}.  Using Eq.~\eqref{eq:cons} repeatedly, we find
\begin{equation}
[i\hat{H},\hat{P}]=-\hat{j}_{L+\frac{1}{2}}+\frac{1}{L}\sum_{x=1}^L\hat{j}_{x-\frac{1}{2}}.
\end{equation}
Because the boundary is open and the current cannot leak out from or flow into the system, we have $\hat{j}_{L+\frac{1}{2}}=\hat{j}_{\frac{1}{2}}=0$ so that
\begin{equation}
[i\hat{H},\hat{P}]=\hat{\bar{j}},
\end{equation}
where $\hat{\bar{j}}$ is the averaged current operator given in Eq.~\eqref{eq:ac}. This is the many-body version of the relation $\hat{v}=i[\hat{H},\hat{x}]$ in the single-particle quantum mechanics.  Eq.~\eqref{openbc} follows immediately by combining this expression of $\hat{\bar{j}}$ with $\hat{H}|\text{GS}\rangle=E_{\text{GS}}|\text{GS}\rangle$ and $[\hat{H},\hat{\rho}_0]=0$.

\section{Higher dimensions}
\label{sec:highd}
Models in higher dimensions can be treated by reducing them to one dimension either by compactifying all other directions by the periodic boundary condition or by taking a finite-width strip with the open boundary condition \cite{PhysRevD.92.085011,Tada2016,Bachmann,PhysRevB.37.6050}.  All quantities then contain an additional summation over transverse directions. We still impose the periodic boundary condition in $x$. 
Below we provide more details for two spatial dimensions.

\subsection{Two dimensions}
\begin{figure}[t]
\includegraphics[width=1.0\columnwidth]{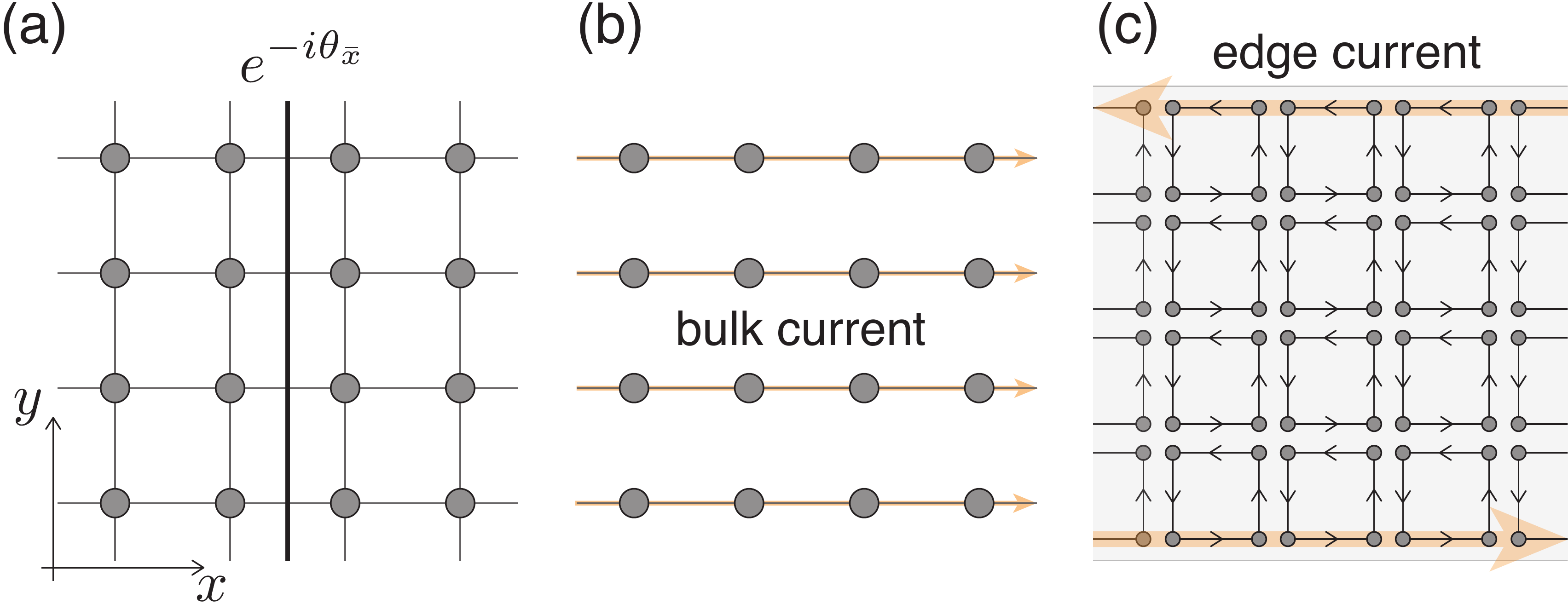}
\caption{\label{fig3} (a) Two dimensional counterpart of Fig.~\ref{fig}. (b) Example of an insulator with a nonzero bulk current. (c) Example of an insulator with a nonzero edge current.
}\end{figure}

We define the current operator in the same way as we did in Sec.~\ref{subsec:current}.
When the twisted boundary condition is introduced at $\bar{x}$, all terms in the Hamiltonian across the seam [the black line in Fig.~\ref{fig3} (a)] acquires the phase $e^{i\ell_{\bar{x}}\theta_{\bar{x}}}$ as before.
\begin{equation}
\partial_{\theta_{\bar{x}}}\hat{H}^{\theta_{\bar{x}}}\big|_{\theta_{\bar{x}}=0}=\sum_{y=1}^{L_y}\hat{j}_{(\bar{x},y)}^x.
\end{equation}
Correspondingly, Eqs.~\eqref{eq:ac}, \eqref{twistop} and \eqref{eq:Taylor} become
\begin{align}
&\hat{\bar{j}}^x\equiv\frac{1}{L_xL_y}\sum_{x=1}^{L_x}\sum_{y=1}^{L_y}\hat{j}_{(\bar{x},y)}^x,\\
&\hat{U}_m\equiv e^{\frac{2\pi im}{L_x}\sum_{x=1}^{L_x}\sum_{y=1}^{L_y}x\hat{n}_{(x,y)}},\\
&\hat{U}_m^\dagger\hat{H}\hat{U}_m=\hat{H}+2\pi m L_y\hat{\bar{j}}^x+O(L_x^{-1}L_y).
\end{align}
In this case, the Bloch theorem states that the expectation value of the averaged current density vanishes in the ground state or in the thermal equilibrium at $T>0$:
\begin{equation}
\lim_{L_x\rightarrow\infty}\langle\text{GS}|\hat{\bar{j}}^x|\text{GS}\rangle=\lim_{L_x\rightarrow\infty}\text{tr}\big(\hat{\rho}_0\hat{\bar{j}}^x\big)=0.
\end{equation}
It is important to note, however, that both of the following quantities may have a non-vanishing expectation value in the limit of large $L_x$. 
\begin{enumerate}
\item The total current integrated over transverse directions ($L_y\hat{\bar{j}}^x$ in two dimensions).
\item The local current density ($\hat{j}_{(\bar{x},y)}^x$ in two dimensions).
\end{enumerate}

\subsection{Bulk current}
An example of the first case is given by the $L_y$ copies of decoupled 1D chains described by the tight-binding model in Sec.~\ref{subsec:persistent}. As each chain supports a persistent current of $O(L_x^{-1})$, we find
\begin{equation}
L_yj_{\text{max}}^x=\tfrac{L_y}{L_x}2t\sin(\tfrac{\pi N}{L_xL_y})=O(L_y/L_x).
\end{equation}
For example, when $L_y=L_x$, this is an $O(1)$ quantity which does not vanish in the large $L_x$ limit.

\subsection{Edge current}
To provide an example of the second case, 
let us consider a two-dimensional periodic array of decoupled 1D rings and impose the open boundary condition in $y$ [Fig.~\ref{fig3} (b)]. Each 1D ring is formed by the tight-binding model considered in Sec.~\ref{subsec:persistent} with $L=4$ and $N=1$. Every ring supports a loop current 
\begin{equation}
j_{\text{loop}}=\tfrac{t}{2}\sin\phi_0,\quad \phi_0\in(-\tfrac{\pi}{4},\tfrac{\pi}{4}).
\end{equation}
This is an $O(1)$ quantity, independent of $L_x$ or $L_y$. In the bulk, contributions from neighboring loops cancel and the local current density vanishes after proper coarse-graining~\cite{jackson1999}. However, at the edge, there is a residual contribution that flows along the edge as illustrated in Fig.~\ref{fig3} (b) implying the nonzero expectation value of $\hat{j}_{(\bar{x},y)}^x$ at the edge. This is nothing but the magnetization current $\vec{\nabla}\times\vec{m}$ originating from the orbital magnetization $\vec{m}$ produced by the loop currents~\cite{jackson1999}.

Of course, a similar situation occurs for Chern insulators but our model is simpler as gapless chiral edge modes are absent.

\section{Conclusion}
In this work we clarified the actual assumption of the Bloch theorem, which was unclear in the discussions in the literature and is turns out to be just the U(1) symmetry and the locality of the Hamiltonian.  We gave a simple proof for the most general version of the theorem.  Since the theorem holds regardless of the details of the states, as far as they are in the ground state or in a thermal equilibrium, the same upper bound on the persistent current applies, for example, to superconductors~\cite{Tada2016}. 

We also clarified the difference of the statement under periodic and open boundary conditions: in the periodic case the current density can be the order of $L^{-1}$, in contrast to the case with the open boundary condition where the current expectation value identically vanishes without taking the thermodynamic limit.  

Finally we discussed a few illuminating tight-binding models in which (i) the net current flow integrated over the transverse direction or (ii) the local current density near the boundary does not vanish even in the limit of the large system size. These non-vanishing current expectation values do not contradict with the general theorem.

\begin{acknowledgements}
I would like to thank S. Bachmann, A. Kapustin, and L. Trifunovic for useful discussions on this topic.  
In particular, I learned the generalization to the Gibbs state from the authors of Ref.~\cite{Bachmann} and I am indebted to their email correspondence.
I also thank T. Momoi for informing us of Ref.~\cite{PhysRevB.78.144404}.
This work is supported by JST PRESTO Grant No.~JPMJPR18LA. 
\end{acknowledgements}

\end{document}